\documentclass{sigchi}

 \toappear{
 Permission to make digital or hard copies of all or part of this work for personal or classroom use is granted without fee provided that copies are not made or distributed for profit or commercial advantage and that copies bear this notice and the full citation on the first page. Copyrights for components of this work owned by others than the author(s) must be honored. Abstracting with credit is permitted. To copy otherwise, or republish, to post on servers or to redistribute to lists, requires prior specific permission and/or a fee. Request permissions from \href{mailto:Permissions@acm.org}{Permissions@acm.org}. \\
 \emph{CHI PLAY '18}, October 28--31, 2018, Melbourne, VIC, Australia \\
 \textcopyright~2018 Copyright is held by the owner/author(s). Publication rights licensed to ACM. 
 This is the author’s version of the work. It is posted here for your personal use. Not for redistribution.
 \\
 ACM 978-1-4503-5624-4/18/10\ldots \$15.00 \\
 DOI: \url{https://dx.doi.org/10.1145/3242671.3242704}
 }

\usepackage{balance}       
\usepackage{graphics}      
\usepackage[T1]{fontenc}   
\usepackage{txfonts}
\usepackage{mathptmx}
\usepackage[pdflang={en-US},pdftex]{hyperref}
\usepackage{color}
\usepackage{booktabs}
\usepackage{textcomp}

\usepackage{microtype}        
\usepackage{ccicons}          

\usepackage{todonotes}

\def\plaintitle{GulliVR: A Walking-Oriented Technique for Navigation in Virtual Reality Games Based on Virtual Body Resizing}

\def\emptyauthor{}
\def\plainkeywords{Virtual reality games; navigation; physical walking; presence; virtual body size; miniature world.}

\makeatletter
\def\url@leostyle{%
  \@ifundefined{selectfont}{
    \def\UrlFont{\sf}
  }{
    \def\UrlFont{\small\bf\ttfamily}
  }}
\makeatother
\def\pprw{8.5in}
\def\pprh{11in}

\definecolor{linkColor}{RGB}{6,125,233}
\hypersetup{%
  pdftitle={\plaintitle},
  pdfauthor={\emptyauthor},
  pdfkeywords={\plainkeywords},
  pdfdisplaydoctitle=true, 
  bookmarksnumbered,
  pdfstartview={FitH},
  colorlinks,
  citecolor=black,
  filecolor=black,
  linkcolor=black,
  urlcolor=linkColor,
  breaklinks=true,
  hypertexnames=false
}

\teaser{
  \includegraphics[width=\linewidth]{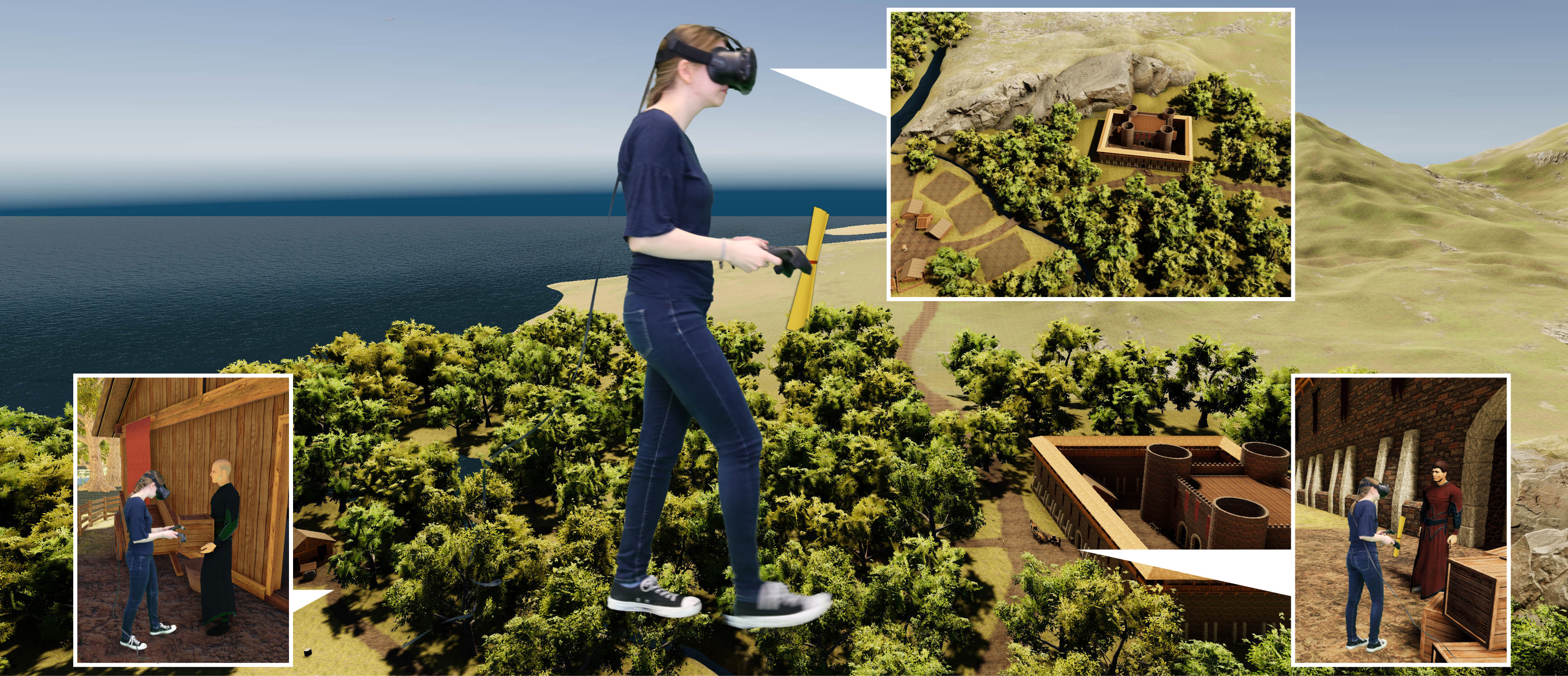}
  
  \caption{GulliVR allows players to become giants on demand and to traverse larger distances in room-scale VR setups within a few steps. The  modeled eye distance is enlarged proportionally to the virtual body size, which ensures the absence of cybersickness.}
  \label{fig:teaser}
}

\usepackage{microtype}

\usepackage{ulem}
\makeatletter
\def\ifEmpty#1{\def\@temp{#1}\ifx\@temp\@empty}
\makeatother

\usepackage{xspace}

\usepackage{amsmath,amsfonts,amssymb}
\usepackage{url} 

\newcommand{\FG}[1]{Figure~\ref{#1}}

\newcommand{\shah}{{\textstyle \amalg{\kern-4.pt\amalg}}}

\usepackage{dblfloatfix}
\begin{document}

\title{\plaintitle}

  \numberofauthors{1}
\author{
  \alignauthor Andrey Krekhov$^{1}$, Sebastian Cmentowski$^{1}$, Katharina Emmerich$^{2}$, Maic Masuch$^{2}$, Jens Kr\"uger$^{1}$ \\
    \affaddr{$^{1}$High Performance Computing Group, $^{2}$Entertainment Computing Group}\\
    \affaddr{University of Duisburg-Essen, Germany}\\
    \email{\{andrey.krekhov, sebastian.cmentowski, katharina.emmerich, maic.masuch, jens.krueger\}@uni-due.de}\\
  }

\maketitle

\begin{abstract}
Virtual reality games are often centered around our feeling of ``being there''. That presence can be significantly enhanced by supporting physical walking. Although modern virtual reality systems enable room-scale motions, the size of our living rooms is not enough to explore vast virtual environments. Developers bypass that limitation by adding virtual navigation such as teleportation. Although such techniques are intended (or designed) to extend but not replace natural walking, what we often observe are nonmoving players beaming to a location that is one real step ahead. Our navigation metaphor emphasizes physical walking by promoting players into giants on demand to cover large distances. In contrast to flying, our technique proportionally increases the modeled eye distance, preventing cybersickness and creating the feeling of being in a miniature world. Our evaluations underpin a significantly increased presence and walking distance compared to the teleportation approach. Finally, we derive a set of game design implications related to the integration of our technique.

\end{abstract}




\section{Introduction}

The number of players who discover virtual reality (VR) games for themselves is steadily increasing. Players enjoy the experienced presence in varying virtual worlds, and game developers attempt to design player interaction to be as natural as possible to further enhance the players' feeling of being there. 

In particular, room-scale systems offer the most natural kind of navigation for games--physical walking. Although researchers have emphasized the  superiority of that ``technique'' over approaches such as walking in place or flying~\cite{usoh1999walking} regarding presence, the size of our living rooms imposes a significant limitation that needs to be bypassed. One of the most prominent remedies promoted by VR systems is the teleport method, which has been developed as addition to common walking to overcome the room size restriction. However, in reality, we observe that players hardly move at all. Instead, they teleport to a location that may be only one step ahead.

Our research attempts to increase the amount of player movement and the experienced presence by introducing a novel natural walking navigation technique. Inspired by \textit{Gulliver's Travels}~\cite{swift1995gulliver}, the idea behind our \textit{GulliVR} approach is to enlarge the virtual body of the player on demand to allow travel over large distances within a few steps, as depicted in \FG{fig:teaser}. In the giant mode, we proportionally increase the modeled eye distance to keep the physical movement and the perceived visual feedback in sync, which prevents cybersickness and creates the impression of walking through a miniature world.

Scaling for navigation purposes is not novel in VR and is often employed in world-in-miniature~\cite{stoakley1995virtual} techniques that rely on a demagnified version of the world. Scaling the user is known from multi-scale VR applications~\cite{argelaguet2016giant} and is closely related to our method. The key difference and our first contribution is that we explicitly modify the depth perception by adjusting the modeled eye separation on the fly to match the player size and, thus, to prevent cybersickness. As a byproduct, the altered depth perception in the enlarged mode leads to an impression of navigating through a toy world.

Our second contribution is the application of navigation-related player rescaling to the domain of VR games. We demonstrate a possible embedding of our technique in a 3D adventure and compare our method to the state of the art teleportation approach. The study reveals a significant increase in presence when using GulliVR, which is a valuable reason for game developers to consider our method as a game mechanic. GulliVR is not a universal remedy for the VR navigation problem, but rather an alternative locomotion technique with its unique strengths and weaknesses. Therefore, our paper also exposes several game design implications that are meant to help developers to decide whether and how the method should be considered for a particular purpose.

\section{Related Work}

Our work belongs to gaming-oriented VR locomotion research. We briefly introduce basic concepts of playing games in VR before discussing in depth the fundamental research on VR locomotion. As GulliVR is centered around the variation of pupillary distance and size perception phenomena, we also provide background related to these topics to explain the mechanisms behind our method.

\subsection{Player Experience in Virtual Reality}

Player experience is one of the most utilized and well-known concepts in our community and consists of multiple components, such as competence, challenge, immersion, and flow. Consequently, various established methods exist that allow measurement of the player experience, such as the evaluation approaches presented by Bernhaupt~\cite{Bernhaupt2010}. These approaches were further formalized by IJsselsteijn et al.~\cite{ijsselsteijn2007characterising, ijsselsteijn2008measuring} and sublimed in the Game Experience Questionnaire (GEQ)~\cite{IJsselsteijn.2013}. Regarding the overall feelings and experiences of players, Poels et al.~\cite{Poels:2007:ALF:1328202.1328218} summarized a game experience categorization based on a focus group study.

When addressing player experience in a virtual environment, immersion~\cite{cairns2014immersion} often plays a leading role. The term \textit{immersion} is mainly used to describe the technical quality of a VR setup~\cite{Biocca:1995:IVR:207922.207926, sherman2002understanding}. If we want to describe how immersive tools impact our perception, the term \textit{presence} is usually chosen by researchers. For an in-depth description and formalization of immersion and presence, we point to the work by Slater et al.~\cite{slater1995taking, slater2003note}. The former paper~\cite{slater1995taking} also illustrates how locomotion in particular impacts presence. Further definitions and measurement techniques related to presence, also referred to as the feeling of being there~\cite{heeter1992being}, can be found, e.g., in the work of Lombard et al.~\cite{lombard1997heart} and IJsselsteijn et al.~\cite{IJsselsteijn}.

Navigating in VR comes with the well-known risk of \textit{cybersickness}~\cite{laviola2000discussion}. Before focusing on that phenomenon, we mention two similar issues: motion sickness~\cite{money1970motion} and simulator sickness~\cite{kolasinski1995simulator}. All three types are similar in their symptoms such as headache, eye strain, sweating, nausea, and vomiting. However, the cause usually differs. Motion sickness happens when our inner ear senses a movement that does not correspond to our visually perceived movement~\cite{reason1975motion}. Note that motion sickness is sometimes used in VR context~\cite{ohyama2007autonomic, hettinger1992visually}. Simulator sickness usually happens when a simulator, such as used for pilot training, does not exactly reproduce the visual movement~\cite{kennedy1989simulator}. That term is also commonly used in VR context, and researchers often rely on the Simulator Sickness Questionnaire (SSQ)~\cite{kennedy1993simulator} to evaluate a VR experience. The work by Stanney et al.~\cite{stanney1997cybersickness} points out that these sicknesses are not the same and and are characterized by different predominant symptoms. Regarding the different reasons for cybersickness, we point the readers to the discussion by LaViola Jr~\cite{laviola2000discussion}. Apart from the obvious technological issues such as flickering and lags, the author also provides a summary of  the most prominent theories about cybersickness: sensory conflict theory (most accepted), poison theory, and postural instability theory. Applying the sensory conflict theory means that moving in VR should trigger an appropriate visual response, and vice versa, if we want to obviate cybersickness.

One possible approach to reduce cybersickness is a reduction of the field of view, as recently proposed by Fernandes et al.~\cite{fernandes2016combating}. A broader discussion about the influence of the field of view on the VR experience can be found in the work of Lin et al.~\cite{lin2002effects}. In the area of digital games, von Mammen et al.~\cite{von2016cyber} conducted experiments by deliberately inducing cybersickness in VR game participants and determined that such games could still be enjoyable. Iskenderova et al.~\cite{iskenderova2017drunk} went a step further by by providing the subjects with alcohol to see how it would influence the symptoms of cybersickness. Surprisingly, these symptoms were significantly reduced at a moderate blood alcohol level around 0.07\%.

\subsection{Locomotion}

Designing navigation in VR games is a challenging task~\cite{Habgood:2017:HLP:3130859.3131437}. Most of the non-VR navigation techniques rely on joysticks and keyboards and induce static player poses, which reduces  presence and leads to cybersickness due to sensory conflicts. In contrast, natural walking~\cite{ruddle2009benefits} or even just walking in place~\cite{slater1995taking, tregillus2016vr} can significantly increase presence. Bhandari et al.~\cite{Bhandari:2017:LSW:3139131.3139133} combined both walking approaches in their \textit{Legomotion} algorithm and reported a higher presence compared to controller input. Another comparison was presented by Usoh et al.~\cite{usoh1999walking}: according to the authors, walking outperforms walking in place, and both are superior to flying regarding presence. That finding was another reason for us to establish the walking metaphor as a giant rather than flying as a bird. Natural walking is also more efficient compared to virtual travel when the navigation task resembles real-world behavior~\cite{suma2010evaluation}. Furthermore, Ruddle et al.~\cite{ruddle2011walking} demonstrated that walking positively influences the cognitive map in large environments.

The current state of the art regarding locomotion in VR games was recently summarized by Habgood et al.~\cite{Habgood:2017:HLP:3130859.3131437}. One conclusion drawn by the authors is ``that short, fast movements in VR (with no acceleration or deceleration) don't appear to induce significant feelings of motion sickness for most users''. Our prestudies confirm that finding, as the transformation into the giant mode was perceived most comfortable when carried out very quickly or instantly. An additional backup regarding fast movements can be found in the work of Medeiros et al.~\cite{medeiros2016effects} and the guidelines by Yao et al.~\cite{yao2014oculus}. The same positive effect holds for the arc-based teleportation technique that is promoted and encouraged by the majority of established VR systems such as the HTC Vive~\cite{vive}. 

The major issue of natural walking is the limited space, as players can take only a few steps in each direction. Hence, researchers are constantly attempting to overcome that obstacle by introducing novel navigation metaphors that go beyond the previously mentioned teleportation technique. For example, Interrante et al.~\cite{interrante2007seven} proposed \textit{Seven League Boots}. The technique deduces the intended travel direction and augments the corresponding component while leaving other directions unscaled, allowing the user to travel forward at increased speed. Steinicke et al.~\cite{steinicke2007hybrid} focused on geospatial environments as an application domain for VR where the virtual space clearly exceeds the tracked area. The authors suggested several hybrid approaches such as the rocket-belt metaphor as an alternative flying approach and visual bookmarks that allow quick jumps to previously marked locations of interest.

Certain locomotion techniques have not found their way into VR games due to their specific requirements. One example is the manifestation of the virtual treadmill concepts~\cite{slater1995virtual} in physical treadmills~\cite{Darken:1997:OTL:263407.263550, omni}. Another important research direction is redirected walking~\cite{razzaque2001redirected, razzaque2005redirected}. Such approaches imperceptibly rotate the virtual environment and force users to slightly change their walking direction. Users think that they are moving straight forward, whereas in reality, they are walking in a large circle. Due to the required space, redirected walking is rarely used for living room sized VR setups, and requires further adaptations~\cite{Grechkin:2016:RDT:2931002.2931018, engel2008psychophysically, langbehn2016subliminal,bruder2009arch} to overcome that limitation.

\subsection{Size and Distance in VR}

GulliVR changes the size ratio between the virtual body and the virtual world. Hence, in the broader sense, our approach can be classified as a multiscale virtual environment (MSVE) navigation. In that context, the technique that most resembles GulliVR is \textit{GiAnt} by Argelaguet et al.~\cite{argelaguet2016giant}, but with the focus on automated speed and scale factor adjustments to account for negative effects such as diplopia~\cite{lambooij2009visual}. Similar to our findings, the authors emphasize the advantages of providing a navigation speed that is perceived to be constant by users.

Other experiments on MSVE navigation techniques centered around the exploration of the human body were carried out by Kopper et al.~\cite{kopper2006design}, who confirmed that automatic scaling outperforms user-defined manual scaling. We incorporated these results in our experiment by predefining the virtual body size based on current game objectives. In contrast to the previously mentioned full-body scaling approaches, the Go-Go technique~\cite{poupyrev1996go} changes only the size and reach of the virtual arm to allow the direct manipulation of distant objects.

LaViola Jr et al.~\cite{laviola2001hands} explored hands-free navigation possibilities in MSVE. Their step world-in-miniature (WIM~\cite{stoakley1995virtual}) widget, being a walkable mini-map, resembles our navigation as a giant over the miniature world. Similarly, Valkov et al.~\cite{valkov2010traveling} introduced a combination of multitouch hand gestures and foot gestures to explore a WIM. To increase navigation precision, the preliminary work by Elvezio et al.~\cite{7892386} proposes to control the posture of our avatar after WIM-triggered travel by post-teleport previews. As suggested by Bruder et al.~\cite{bruder2009arch}, such previews can also be used as virtual portals where users have to pass through to reach the displayed destination. The portal concept works especially well when such ``doors'' can be naturally integrated into the virtual scenario.

One issue of traditional WIM approaches is the lack of adaptation to differently scaled virtual worlds. To overcome that limitation, Wingrave et al.~\cite{wingrave2006overcoming} proposed a scaled scrolling world-in-miniature (SSWIM). SSWIM allows users to zoom and scroll the miniature representation of the world, which simplifies navigation and interaction in cases when the world is very large or small.

An important question for game design using the GulliVR technique is whether and how such scale adjustments influence our perception and interaction in VR. In general, researchers agree that we usually underestimate distance in VR~\cite{frenz2007estimation, renner2013perception, cutting1995perceiving}. Although object size familiarity plays an important role for distance experiment~\cite{nguyen2011effects, ogawa2017distortion}, the sense of our own body is a major factor regarding our judgments on objects' size and distance~\cite{kokkinara2015effects}. The body effect was also extensively studied by van der Hoort et al.~\cite{van2011being}, who concluded that a user in a large virtual body perceives the objects smaller and nearer. The effect perfectly aligns with our experiments: our subjects, as giants, reported seeing the environment as a miniature toy world. Although we do not display any body parts in our testbed game, the research by Jun et al.~\cite{jun2015big} might be an interesting starting point regarding the players' mental ability to step over virtual obstacles in a game, because the authors found that displaying large feet would allow the user to step over larger gaps.

A major difference between flying navigation and GulliVR is the increased interpupillary distance (IPD) of the latter as a result of enlarging the whole virtual body. Although tiny variations of IPD have been shown to have no measurable impact on size judgments~\cite{best1996perceptual} and can even be applied unnoticed~\cite{ware1998dynamic}, setting the modeled eye separation to a significantly different value compared to the physical eye separation results in so-called \textit{false eye separation}~\cite{cho2014evaluating}. The resulting perceived image causes an altered size perception compared to real objects~\cite{wartell1999analytic}, leading to our miniature world perspective. Similar findings were also reported by Renner et al.~\cite{renner2015influence}, confirming that increasing the stereo base (i.e., the modeled eye distance) makes objects appear nearer and smaller.

\begin{figure}
\centering
\includegraphics[width=1.0\columnwidth]{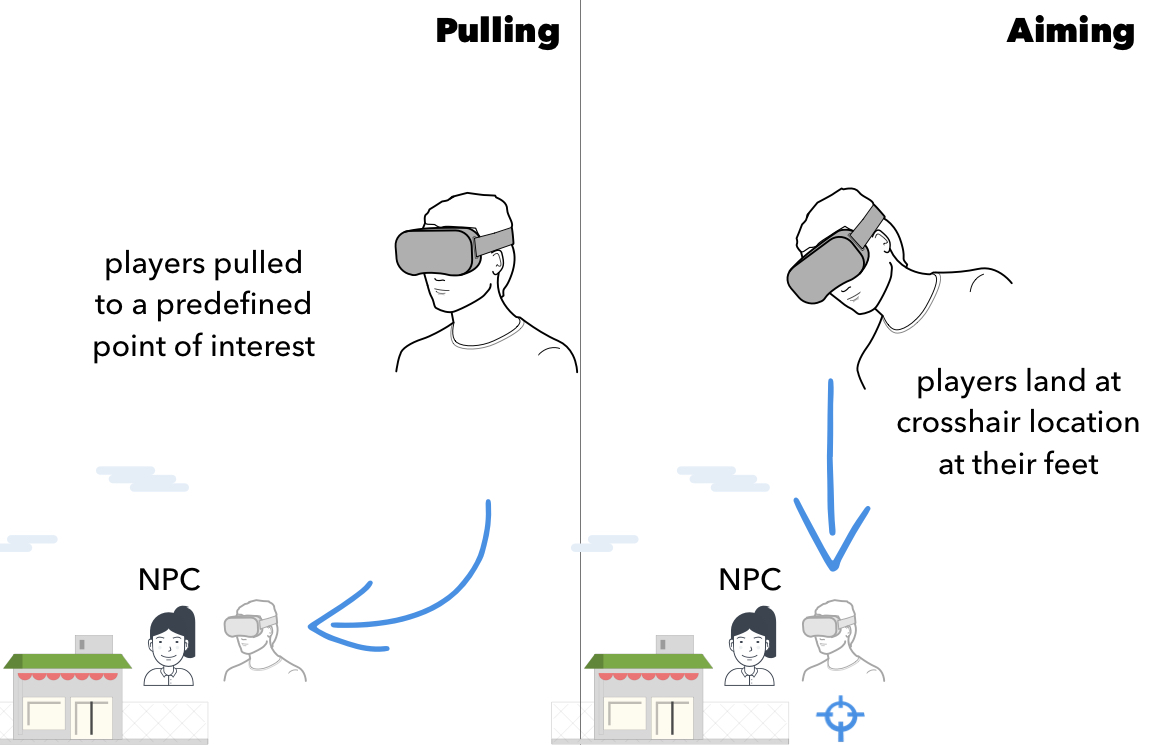}
\caption{\textit{Pulling} (left) and \textit{aiming} (right) are two possibilities that allow precise transitions from GM to NM. Pulling adds a horizontal translation toward the nearby point of interest, whereas aiming displays a crosshair next to players' feet to indicate the destination location.}
\label{fig:aiming}
\vskip -1em
\end{figure}

\section{GulliVR Navigation}

The main idea behind GulliVR is to enlarge the virtual body of the player on demand. This \textit{giant mode (GM)} allows the player to travel large distances in a room-scale environment using natural walking. Once the player reaches his or her destination, the virtual body size is reset to \textit{normal mode (NM)}. Naturally, we can obtain the same results by shrinking the size of the world instead of enlarging the player. However, we would not recommend that approach in practice for performance reasons. For instance, having a fully resizable environment usually interferes with baked lighting.

\begin{figure}
\centering
\includegraphics[width=1.0\columnwidth]{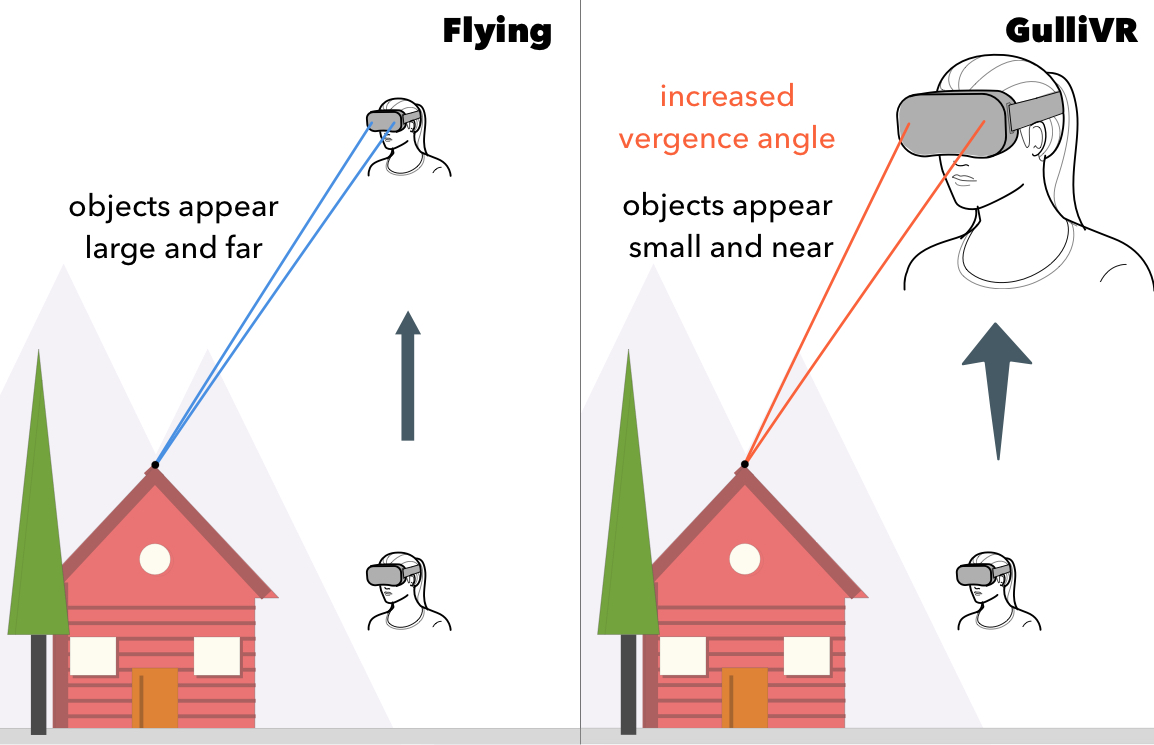}
\caption{Significantly increasing the modeled eye distance results in a larger vergence angle, altering the size/distance perception of objects and evoking the feeling of being a giant. Physical movement speed is perfectly aligned with the visual feedback, which obviates cybersickness.}
\label{fig:eyes}
\end{figure}

\begin{figure*}[t!]
\centering
\includegraphics[width=2.1\columnwidth]{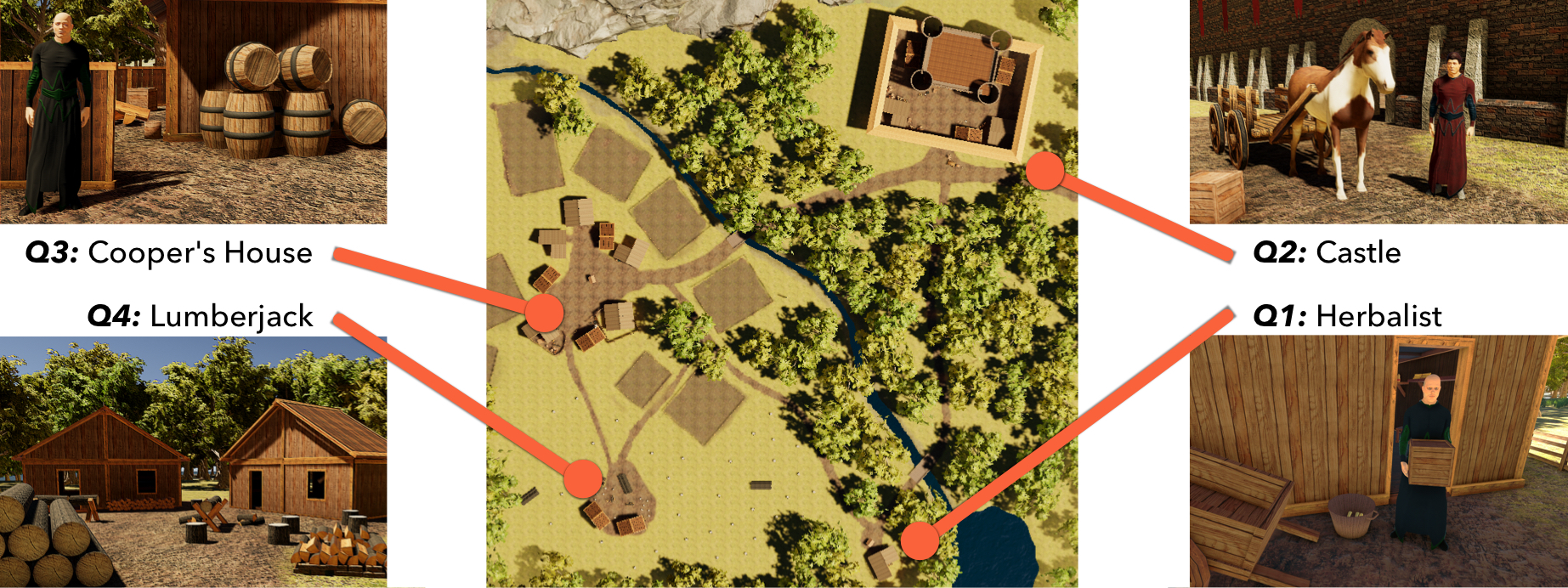}
\caption{Overview of the virtual testbed game, including points of interest and related quests.}
\label{fig:game}
\end{figure*}

Players need to control their transition from GM to NM, because ``landing'' somewhere offside is frustrating. We outline two approaches: active \textit{aiming} and passive \textit{pulling}, as depicted in \FG{fig:aiming}. Aiming can be realized by displaying crosshairs when the player looks down. Pulling is what we utilized for our testbed game. In that case, certain points of interest can be enhanced with bounding boxes. If the player initiates the transition to NM inside such a box, the virtual body is pulled toward the predefined interesting position, e.g., in front of an NPC. Triggering a transition outside the bounding box can be either prohibited or behave just like active aiming, with or without crosshairs. We assume that active aiming is more suited for games with free exploration modes, whereas pulling is useful for strictly scripted narrative flows.

At this point, we strongly emphasize the difference between enlarging the virtual body and flying, i.e., increasing the camera height and scaling its movement speed. In a nonstereo setup, both approaches would work identically. However, in a VR setup, players instantly note the difference. Enlarging the virtual body also means substantially increasing the modeled eye separation, as shown in \FG{fig:eyes}. Hence, players' size and distance perception of objects is altered~\cite{van2011being, renner2015influence} and the world appears in miniature. This alteration has the important benefit that our physical walking speed is perfectly aligned with the visual feedback we receive. Flying, in contrast, usually promotes a cognitive mismatch. Hence, we assume that GulliVR does not induce cybersickness.

Cleary, GulliVR is not a universal remedy for all VR games, and we suppose that readers already have certain counterexamples, e.g., closed environments with ceilings, where the technique would perform poorly. Also note that GulliVR has notable degrees of freedom that developers should be aware of, e.g., GM size, object manipulation, and NPC interaction. We postpone a detailed discussion of the related design space, including possible limitations, until the section \textit{``Design Implications''} to allow us to incorporate the results gathered from our experiments.

\section{Evaluation}

We created a 3D adventure in VR to validate our technique and to explore its drawbacks. In our experiment, we compared a group of players relying on the common arc teleportation method to a group of subjects using GulliVR and measured aspects such as presence and physical walking distances. In addition to this game experiment, we asked the participants to perform several targeting tasks with GulliVR to see how precisely they could tell the system where to arrive after transitioning back to NM.

\subsection{Hypotheses}

The main assumption regarding GulliVR is that emphasizing natural walking should have a positive influence on the players' presence perception. Furthermore, we assume that our manipulations with the modeled eye distance successfully prevent cybersickness, because there is no cognitive mismatch due to the alignment of visual feedback with physical movement. Finally, we hypothesize that players walk significantly more in GulliVR mode compared to the teleport technique. Although teleporting does not prohibit walking, our previous observations indicate that players stay rather still, even if the target is very close. To summarize, our three main hypotheses are:

\begin{itemize}
  \setlength{\itemsep}{2pt}
  \setlength{\parskip}{0pt}
  \setlength{\parsep}{0pt}
\item H1: GulliVR \textbf{significantly increases players' presence} compared to the teleport technique.
\item H2: There is \textbf{no cybersickness} in the GulliVR mode.
\item H3: In comparison to the teleport technique, \textbf{players walk significantly more} when using GulliVR.
\end{itemize}

\subsection{Testbed Scenario}

Our testbed adventure game is realized with the Unity 3D Engine~\cite{unity} and takes place in a medieval, fictive world. The main character is an herbalist's apprentice in a rural area next to a famous castle. In our scenario, the player experiences one day in the life of the apprentice, solves simple quests, and explores the surrounding points of interest.

The world map is depicted in \FG{fig:game}. The player's journey begins in front of the lonely forest house of the herbalist. The herbalist asks the player to take the crate he is holding and to put it on a nearby cart (\textit{\textbf{Q1}}). This first interaction allows the subjects to get familiar with object manipulation by using the trigger button of the controller.

Upon successful completion of the first task, the herbalist tells the player to pick up a healing scroll from the basket and to take it to the diseased lord of a nearby castle (\textit{\textbf{Q2}}). The player also receives a detailed description to get to the castle. As a hint, the herbalist recommends the player make use of the navigation mode, i.e., teleport or GulliVR, for traversing larger distances by pressing the thumb button.

After a while, the player arrives at the castle and encounters a servant, already awaiting the scroll. Handing the scroll over finishes \textit{\textbf{Q2}}, and the servant asks the player to load supply crates (\textit{\textbf{Q3}}) on a cart while he delivers the scroll to the lord.

Loading at least three crates triggers the servant to come back and tell the player to 
revisit the herbalist (\textit{\textbf{Q3}}), who is in a nearby village by now. The herbalist waits in front of the cooper's house and requests the player's help one last time. The cooper needs some logs, and the player is asked to fetch them from the lumberjack (\textit{\textbf{Q4}}). The quest is completed when the player delivers at least four logs. At this point, the scripted part of the game is completed. In an additional part of the game, the player can freely explore the world. A few changes are triggered. First, most of the game objects such as barrels become interactive. Second, a number of Easter eggs are added, e.g., the herbalist's house contains now a candle that, upon being picked up, shifts the game into night mode.

\subsubsection{GulliVR Configuration}

In GulliVR mode, the transformation between GM and NM takes $0.5$ seconds. We chose that time based on a pretest ($N = 5$), because we had determined that any time slower than one second causes serious cybersickness, which is aligned with previous research~\cite{Habgood:2017:HLP:3130859.3131437}. We also preferred fast transformation over instant switching to prevent possible disorientation.

We chose the passive pulling approach for the transition from GM to NM. Quest-related areas, i.e., the castle, the cooper's house, and the lumberjack's house, are enhanced by invisible bounding boxes. Standing inside these areas and triggering the transition pulls the player toward the intended position, e.g., in front of the herbalist. This additional horizontal vector during the downward transition is imperceptible due to the very short transformation time. Going from GM to NM outside the quest-related bounding boxes shrinks the players, but does not pull them in any direction.

The scaling factor for the virtual body is automatically selected with respect to the current quest. Before the completion of \textit{\textbf{Q3}}, the GM is roughly $100x$ larger than NM, since the locations are further away. For \textit{\textbf{Q4}}, i.e., fetching logs, the destination is much closer, and we chose $30x$ as the scaling factor (cf. \FG{fig:gulli}).

\subsection{Procedure and Applied Measures}

We conducted a between-subject experiment with the navigation mode (cf. \FG{fig:gulli}), i.e., teleport and GulliVR, as the independent variable. The main reason for choosing a between-subject design was to minimize sequence effects from repeating quests and from possible cybersickness symptoms. On average, the study took 45 minutes and was conducted in our VR research lab equipped with an HTC Vive. After informing subjects about the study's procedure, we administered a questionnaire to assess age, gender, digital gaming behavior, and prior experiences with VR systems. 
Furthermore, as a control variable, we used the Immersive Tendencies Questionnaire (ITQ) \cite{Witmer.1998} to assess how easily participants get immersed in activities like gaming and watching movies. The ITQ includes the four subscales \textit{involvement}, \textit{focus}, \textit{games}, and \textit{emotions}.
We then introduced the subjects to the HTC Vive and explained the game controls: the trigger button is for picking and holding objects, and the thumb button activates the navigation mode.

In the teleport case, we explicitly explained how the teleportation works. In the GulliVR condition, we mentioned only that the thumb button triggers a navigation mechanism that is helpful for the traversal of larger distances. We intentionally neither embedded the technique in the story context nor provided any relational cues that could bias the subjects' impression of being a giant in order to gather reliable knowledge on how clueless players would perceive such a perspective.

After the briefing, subjects played the first part of the game (\textit{\textbf{Q1}}-\textit{\textbf{Q4}}). We logged all quest durations, interactions, and the physical distance traveled by the subjects. Upon completion of the final quest, an in-game message indicated a pause and advised the participants to remove the head-mounted display.

Subsequently, we administered questionnaires related to presence, player experience, and cybersickness. To assess feelings of presence, we relied on the Igroup Presence Questionnaire (IPQ) \cite{Schubert.1999b}. The IPQ contains the three subdimensions \textit{spatial presence}, \textit{involvement}, and \textit{experienced realism}, as well as one single item to assess perceived general presence (\textit{``In the computer generated world, I had a sense of `being there' ''}). All items are phrased as statements that participants have to rate on a 7-point Likert scale (coded 0~-~6).

As we are particularly interested in the influence of our navigation technique on feelings of presence, we additionally administered the Presence Questionnaire (PQ) (originally developed by Witmer and Singer \cite{Witmer.1998} and revised by the UQO Cyberpsychology Lab \cite{UQO.2004}). The PQ includes the subdimensions \textit{realism}, \textit{possibility to act}, \textit{quality of interface}, \textit{possibility to examine}, and \textit{self-evaluation of performance} (coded 0~-~6). By focusing more on the interactions with and navigation through the game environment, the PQ is a good complement of the IPQ to assess all aspects of presence.

To measure further dimensions of the overall player experience, we applied the Game Experience Questionnaire (GEQ)~\cite{IJsselsteijn.2013}, which consists of seven subscales: \textit{positive affect}, \textit{negative affect}, \textit{immersion}, \textit{flow}, \textit{challenge}, \textit{tension/annoyance}, and \textit{competence}. All 33 items are presented in the form of statements to which participants rate their agreement on a 5-point Likert scale (coded 0~-~4). The Simulator Sickness Questionnaire (SSQ) \cite{kennedy1993simulator} was administered to test whether one of our navigation techniques or the game in general causes any negative physical reactions in terms of cybersickness. It assesses symptoms of cybersickness on the three subscales \textit{nausea}, \textit{oculomotor}, and \textit{disorientation} on a rating scale ranging from 0 (none) to 3 (severe).
Finally, we asked some custom questions specifically addressing the navigation technique used for which participants had to rate their agreement on a 7-point Likert scale (see Table~\ref{tab:Custom}).

\begin{figure}
\centering
\includegraphics[width=1.0\columnwidth]{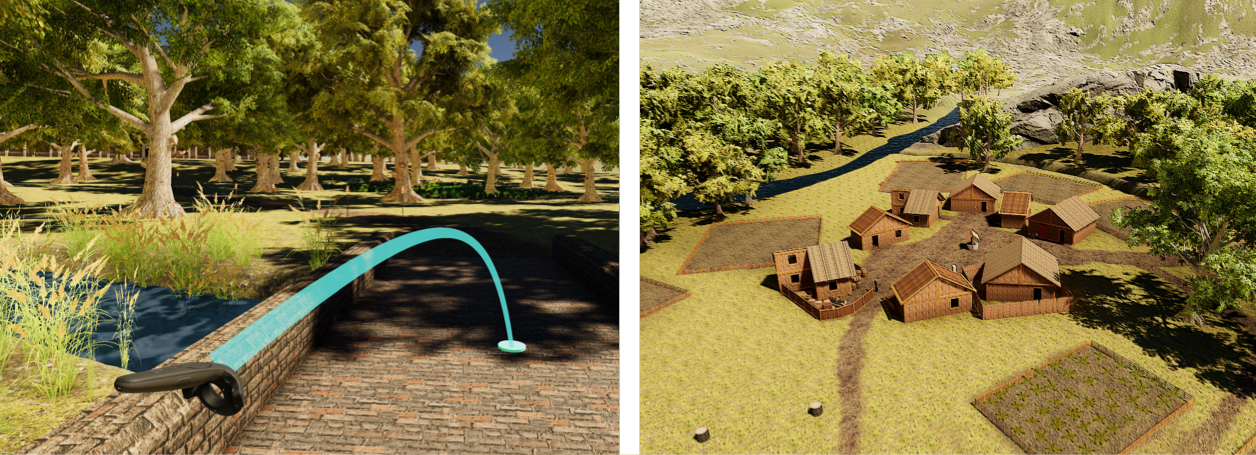}
\caption{Player's perspective when using the arc teleportation technique (left) and in GulliVR mode (right) with 30$x$ virtual body scaling.}
\label{fig:gulli}
\end{figure}

\begin{table*}[t]
  \caption{Mean scores, standard deviations, and independent samples t-test values of the iGroup Presence Questionnaire (IPQ) and the Presence Questionnaire (PQ).}
  \label{tab:IPQ}
  \begin{tabular}{lccccl}
    \toprule
     & ~~~~~~~GulliVR ($N = 15$) & ~~~~~~~Teleportation ($N = 15$) & ~~~~~~~ & ~~~~~~~ &\\
     \addlinespace 
      & ~~~~~~~M (SD)	& ~~~~~~~M (SD) & ~~~~~~~\textit{t} (28) & ~~~~~~~Significance \textit{p} & \\
    \midrule
    IPQ (scale: 0 - 6)\\
    \ \ \ \ \ \ Spatial Presence & ~~~~~~~3.84 (0.73) & ~~~~~~~3.21 (0.93) & ~~~~~~~-2.05 & ~~~~~~~.050 & \\
    \ \ \ \ \ \ Involvement & ~~~~~~~3.42 (0.82) & ~~~~~~~3.07 (1.11) & ~~~~~~~-0.98 & ~~~~~~~.334 &\\
    \ \ \ \ \ \ Realism & ~~~~~~~2.50 (0.75) & ~~~~~~~2.17 (0.63) & ~~~~~~~-1.32 & ~~~~~~~.199 &\\
    \ \ \ \ \ \ General & ~~~~~~~4.40 (1.12) & ~~~~~~~3.00 (1.51) & ~~~~~~~-2.88 & ~~~~~~~.008& **\\
    PQ (scale: 0 - 6)\\
    \ \ \ \ \ \ Realism & ~~~~~~~4.33 (0.84) & ~~~~~~~3.33 (0.63) & ~~~~~~~-3.69 & ~~~~~~~.001 & **\\
    \ \ \ \ \ \ Possibility to Act & ~~~~~~~4.12 (0.94) & ~~~~~~~3.27 (0.79) & ~~~~~~~-2.67 & ~~~~~~~.013& *\\
    \ \ \ \ \ \ Interface Quality & ~~~~~~~5.20 (0.77) & ~~~~~~~5.13 (0.70) & ~~~~~~~-0.25 & ~~~~~~~.806&\\
    \ \ \ \ \ \ Possibility to Examine & ~~~~~~~4.78 (0.81) & ~~~~~~~3.98 (0.88) & ~~~~~~~-2.59 & ~~~~~~~.015 & *\\
    \ \ \ \ \ \ Performance & ~~~~~~~4.70 (0.92) & ~~~~~~~4.67 (1.32) & ~~~~~~~-0.08 & ~~~~~~~.937 &\\
    \ \ \ \ \ \ Total & ~~~~~~~4.53 (0.60) & ~~~~~~~3.85 (0.55) & ~~~~~~~-3.28 & ~~~~~~~.003 & **\\
  \bottomrule
  &&&&*\textit{p} <.05, ** \textit{p} <.01
\end{tabular}
\end{table*}

Upon the completion of the questionnaires, we asked the subjects to return into the virtual reality and freely explore the world. To trigger the players' desire to explore, we mentioned that a certain amount of Easter eggs had been added since their last VR visit. As in the first game round, we logged all interactions, traveled distances, and the overall exploration duration. The subjects were asked to give a signal when they felt that they played enough. Otherwise, we stopped the exploration mode after 20 minutes.

The third part of our experiment consisted of a targeting task and was the same for both condition groups, with the teleport group receiving a brief introduction into the GulliVR technique. We placed the participants into GM ($100x$) and showcased the four target circles as shown in \FG{fig:target}. We asked the subject to attempt to get in the middle of each circle and then switch to NM. After the transformation, the target was colored in accordance with the hit zone. Each target could be visited only twice, and immediate reattempts were not possible. After eight hits, an in-game message indicated the completion of the overall study.

\section{Results}

\subsection{Sample Description}
In sum, 30 persons (15 female) participated in our study with a mean age of 27.6 ($SD=11.18$). All participants reported playing digital games at least a few times a month and the majority of them (26) had also used VR gaming systems before, but they had little experience with such systems. 
Participants were randomly assigned to one of the two study conditions (GulliVR vs. teleportation navigation). 
Participants in the two groups did not differ significantly regarding their distribution of age, gender, and prior experience with VR games. Furthermore, there was no significant difference regarding the immersive tendencies of our subjects ($t=1.47$, $p=.152$).

\subsection{Testing the Hypotheses} 
We compared the results of our measures between the two study conditions. For each analysis, we tested the requirements for parametric calculations (homogeneity of variances and normal distribution of data) with Levene's and Kolmogorov-Smirnov tests. If the requirements were violated, Mann-Whitney U~tests are reported instead of independent samples t-tests for testing significant differences between the groups. 

Concerning H1, we compared the measures of presence between the two study conditions. Table~\ref{tab:IPQ} shows the scores of the IPQ and the results of the calculated independent samples t-tests. For all subdimensions, the GulliVR group scored higher than the group using teleportation. Only the difference regarding the general feeling of presence is significant according to the analysis (\textit{p}~=~.008), although spatial presence is just on the edge of being significant as well (\textit{p}~=~.050).
Table~\ref{tab:IPQ} summarizes the mean scores of the PQ and the results of the t-tests comparing the two groups. Similar to the IPQ scores, the GulliVR group shows a clear tendency to experience higher presence, as mean values are higher for all dimensions. The analysis reveals that the differences in experienced realism, the possibilities to act and examine, and the total presence score are significant. In contrast, the quality of the VR interface and the self-evaluation of performance do not differ significantly.

As stated in H2, we wanted to assure that GulliVR does not negatively affect players in terms of cybersickness. Table~\ref{tab:SSQ} shows the weighted scale scores of the SSQ, which are all rather low.  
With respect to reference values reported by Kennedy et al.~\cite{kennedy1993simulator}, our values indicate that participants had no significant problems with cybersickness in either condition. When comparing the SSQ values of both study groups, participants in the GulliVR condition tended to report fewer symptoms than participants using the teleportation navigation. 
However, according to Mann-Whitney U-tests, these differences are not significant (all \textit{p}~>~.067).

\begin{table}
  \caption{Mean scores and standard deviations of the Simulator Sickness Questionnaire (SSQ).}
  \label{tab:SSQ}
  \begin{tabular}{lcc}
    \toprule
    SSQ Dimension & GulliVR ($N = 15$) & Teleport ($N = 15$)\\
     & M (SD)	& M (SD) \\
    \midrule
    Nausea & 3.82 (6.03) & 14.63 (19.37) \\
    Oculomotor & 10.61 (10.25) & 16.68 (18.83)\\
    Disorientation & 13.92 (20.38) & 22.27 (26.72)\\
    Total & 10.47 (10.21) & 19.95 (21.24)\\
  \bottomrule
\end{tabular}
\end{table}

In H3, we assumed that GulliVR encourages players to walk more, as they have to travel all distances by moving in the real world. Indeed, we observed that participants used the room more extensively in the GulliVR condition. This impression is confirmed by the analysis of the logged gameplay data. We measured the distance that players covered (in the real world) during the story part of the game and in the free exploration phase. As the playing duration was variable, we then calculated the mean score for walked meters per minute to have comparable values. On average, players using the GulliVR technique walked $14.75$ meters per minute ($SD=3.85$), whereas players in the teleportation group walked only $11.38$ meters ($SD=2.61$). A t-test indicates that this difference is highly significant ($t(28)=-2.81$, $p=.009$), supporting our hypothesis. The same effect was found for the free exploration phase: again, players in the GulliVR group walked more ($M=10.92$ $m/min$, $SD=3.06$) than players in the teleportation group ($M=7.30$ $m/min$, $SD=2.22$); $t(28)=-3.72$, $p=.001$. In addition, logged data shows that, on average, players in the GulliVR condition voluntarily played about twice as long in the exploration phase ($M=11.38$ $min$, $SD=4.60$) than participants using the teleportation approach ($M=6.34$ $min$, $SD=2.98$); $t(28)=-3.56$, $p=.001$. During the story-driven first part of the game, the difference in duration was not significant (GulliVR: $M=6.82$ $min$, $SD=2.73$; Teleportation: $M=5.93$ $min$, $SD=1.53$); $t(28)=-1.10$, $p=.280$.

\begin{table*}[]
\caption{Mean scores and standard deviations of the custom questions (CQ) and independent samples t-test values of comparison.}
  \label{tab:Custom}
  \begin{tabular}{llccccl}
    \toprule
     && GulliVR ($N = 15$) & Teleport ($N = 15$) &  &  &\\
     &Question Item & M (SD)	& M (SD) & \textit{t} (28) & Sig. \textit{p} & \\
    \midrule
    CQ1 & I would have preferred to move through\\ 
    &the world using another technique. & 1.20 (1.66) & 2.80 (2.00) & 2.38 & .024 &*\\\addlinespace 
    CQ2& I think I have been walking much in the \\ 
    &(real) room while playing the game. & 4.27 (1.98) & 2.33 (1.76) & -2.83 & .009 &*\\\addlinespace 
    CQ3& While playing, I wondered why I had the \\
    &ability to [teleport/grow and shrink]. & 0.80 (1.57) & 0.27 (0.59) & -1.23 & .228 &\\\addlinespace 
    CQ4& I could orient myself well in the game\\& world. & 4.60 (1.40) & 3.33 (1.29) & -2.57 & .016 & *\\ \addlinespace 
    CQ5& I would have liked to spend more time\\&in the game world. & 5.20 (0.86) & 4.20 (1.37) & -2.39& .024 &*\\ \addlinespace 
        CQ6& From above, the world appeared to me\\& as a miniature or toy world. & 5.27 (0.96) & -- & --& -- &\\
  \bottomrule
  &&&&*\textit{p} <.05, ** \textit{p} <.01
\end{tabular}
\end{table*}

\begin{figure}[!b]
\centering
\includegraphics[width=1.0\columnwidth]{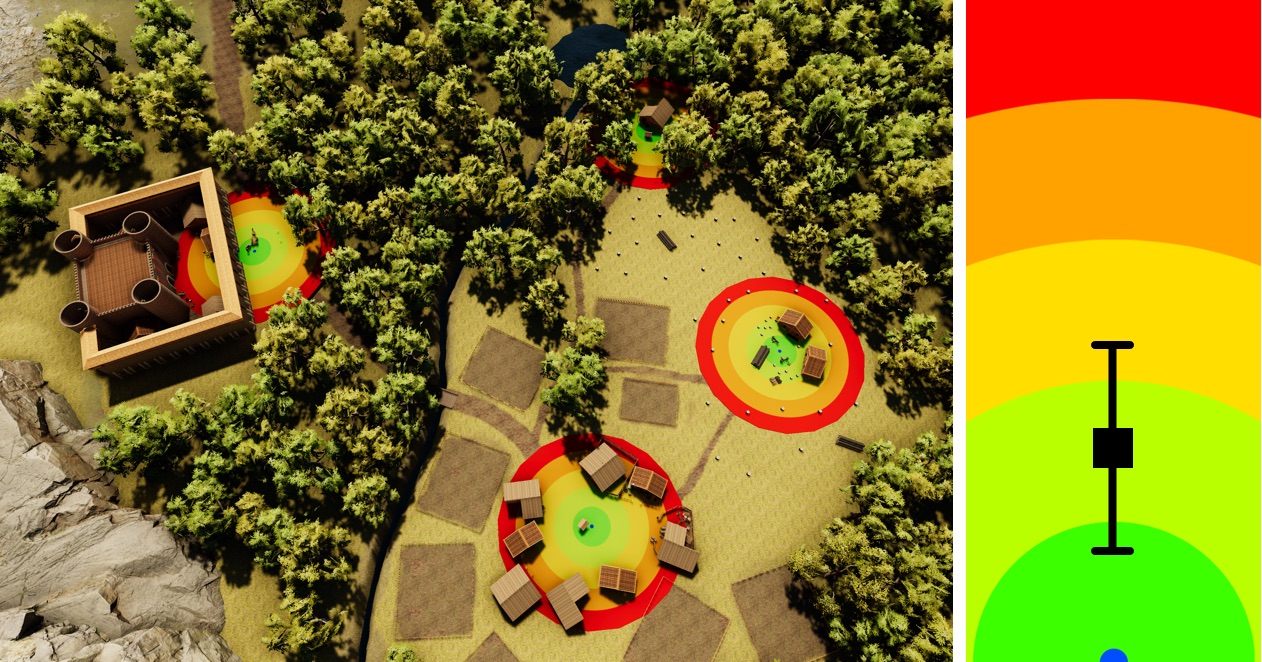}
\caption{The targeting task from GM perspective. The circle has a radius of 0.25 meters in room scale. The illustration on the right exposes the average precision and standard deviation of all participants. }
\label{fig:target}
\end{figure}

\subsection{Further Comparisons Between GulliVR and Teleportation}
Besides testing our hypotheses, our study was aimed at gaining insight into how players experience and evaluate the GulliVR navigation technique, both individually and compared to the established teleportation approach. We compared the ratings of our custom questions and found significant differences between the two groups as outlined in Table~\ref{tab:Custom}.

We were also interested in whether the method of navigation also influences aspects of the player experience other than presence. Hence, we analyzed group differences regarding the seven subdimensions of the GEQ. Results indicate that participants in both groups do not differ significantly regarding any of the subdimensions (all \textit{p}~>~.158).

\subsection{Targeting with GulliVR}

We investigated how precisely players in both groups were able to ``hit'' predefined targets with GulliVR. Results show that, on average, they missed the center of the target by 0.10 meters 
($SD=0.04$) in room-scale metrics. For a comparison, the radius of the target was 0.25 meters, i.e., players ``landed'' closer to the center than to the outer perimeter, as depicted in \FG{fig:target}. 
Our data also indicates a learning effect: players in the GulliVR condition, who had used the technique in the previous phases of the study, performed significantly better in the targeting task ($M=0.07~m$, $SD=0.02$) than players who were completely new to it ($M=0.13~m$, $SD=0.04$); $t(22.29)=3.86$, $p=.001$.

\section{Discussion}

Our results confirm all three hypotheses: GulliVR navigation leads to an increased feeling of presence and makes players walk around in the room frequently without causing cybersickness. The absence of cybersickness promotes our technique to a considerable alternative for navigating a VR game world. In particular, GulliVR induces even slightly less cybersickness compared to the established teleport approach.

Our study demonstrates the major strength of the GulliVR technique: it can significantly increase the players' feelings of presence and thereby supports an intense and positive player experience. Several subscales of our presence measures support that assumption. Notably, the PQ questionnaire reveals that the experience of being able to explore the game world and interact with it is increased. We attribute these results to the fact that the natural conversion of physical steps into in-game navigation allows a more direct relationship with the virtual world. Our finding also aligns with previous research that demonstrated the positive impact of natural walking metaphors~\cite{ruddle2009benefits, slater1995taking, tregillus2016vr, usoh1999walking}. In contrast to GulliVR, teleportation is a rather abstract, and, more importantly, discontinuous approach that potentially increases the disconnection from the virtual world.

One aspect we believe contributes to the positive presence effect is the increased physical activity of players that is necessary to move through the world. The GulliVR technique encourages players to exploit the whole space of the room-scale VR system. Indeed, according to CQ2, subjects in GulliVR mode reported the feeling of walking much in the real room, i.e., players were aware of their increased movement.

Our data also indicates that GulliVR increases players' motivation to explore the game world. First, players spent significantly more time in the voluntary exploration phase. Second, according to CQ5, players reported that they would have liked to spend more time in the game world. Furthermore, participants stated that they were able to orient themselves quite well in the game world, and much better than in the teleport condition (CQ4). Thus, as expected, GulliVR supports orientation and provides a good overview of the landscape/world.

In sum, the GulliVR navigation technique is perceived well by the players, which is also confirmed by the other custom questions. GulliVR participants did not want to have another possibility to navigate through the world (CQ1), whereas this desire was significantly higher in the teleportation group. Interestingly, in neither GulliVR nor teleportation condition did players question the way they were able to move through the game world (CQ3). In other words, although both methods are not very realistic, it is not crucial to explicitly embed these techniques in the game context, because players simply accept them as game mechanics, reminiscent of game twists such as fast travel, which is also not natural but is rarely questioned.

In GM, nearly all participants perceived the virtual environment as a toy world (CQ6). The miniaturization is due to the increased modeled eye distance, and our results confirm and extend previous research in that area~\cite{wartell1999analytic, renner2015influence, van2011being}.

The results also point to one important challenge when using GulliVR, namely precise targeting when switching from GM to NM. Both our observations during the exploration phase and the results of the targeting tasks indicate that players had problems ``landing'' exactly where they wanted to. Although players seem to get used to the aiming approach and performed better over time, we suggest that targeting should be supported by the game system to avoid frustration. Such supportive approaches and further design implications for GulliVR will be discussed in the following section.

\section{Design Implications}
\label{sec:design}

GulliVR extends the toolbox of VR navigation techniques. The section is our preliminary guideline regarding the integration of the technique into VR games and VR applications in general. Hence, we address limitations and important degrees of freedom to be considered when implementing GulliVR.

\subsubsection{Target Acquisition}
Our targeting experiment demonstrated the precision issues of players when switching from GM to NM without any additional help. Therefore, we recommend integrating supportive mechanisms to prevent players from ``landing'' offside. In the section \textit{``GulliVR Navigation''}, we proposed two approaches: passive pulling and active aiming, as also depicted in \FG{fig:aiming}. Our testbed game utilized pulling, as the scenario was rather linear. In that case, players usually did not note that they got pulled to certain locations. However, in the explorative session, players sometimes figured out that they appeared at the same place over and over, although they initiated the transformation to NM from a slightly different place. 

An even more restricting implementation of pulling would be to allow GM to NM transitions only at points of interest. Such as approach can be helpful if players have to traverse larger between-level areas without any game elements. For nonlinear games, we instead recommend the active aiming technique by projecting crosshairs onto the ground next to the player's feet. This technique allows the player to pick the precise destination location at the cost of introducing an additional UI element, i.e., the crosshairs. We assume that such a technique might slightly lower the presence but increase the perceived competence. An experimental validation of that assumption is part of our future work. Note that the proposed technique requires an adaptation for scenarios with multi-level buildings. In those cases, one possibility is to include an additional UI element (floor picker) that explicitly asks the player for the desired level.
\subsubsection{GM Size}
Game designers have the choice between predefined GM size and user-defined manual resizing. According to Kopper et al.~\cite{kopper2006design}, the manual approach is less effective, but might prove otherwise for certain types of games. Recall that in our scenario, we applied predefined GM size, but with dependence on the current game state. Our guideline was to enlarge players to a size where they could travel to the next point of interest within a few steps.

\subsubsection{Resetting}
GulliVR does not completely remove the restrictions imposed by the room size. Hence, game designers should think of a suited resetting mechanism for cases where players reach a physical wall and still want to move forward. GulliVR comes with a built-in reset: switching to NM, making several steps backward, and switching to GM again to overcome the distance in one step. However, such an activity often involves undesired cognitive workload and could be avoided by including other navigation facilities for resetting purposes. For instance, similar to redirected walking, one could rotate the world during NM/GM transitions such that players always move toward the furthermost wall. We propose to evaluate that technique as part of possible future work, as we assume that more studies are needed to, e.g., determine how such rotation would impact the orientation ability of players.

\subsubsection{Modeled Eye Distance}
The virtual eye separation, also called stereo base or modeled eye distance, should be always scaled in proportion to the virtual body size. Otherwise, there is a misalignment between the virtual and the physical floor, resulting in either a feeling of floating/flying (eye distance too low) or standing below the ground (eye distance too high).

\subsubsection{Miniaturization} 
Developers should be aware that increasing the modeled eye distance leads to perceiving the surroundings as a toy world, which could also be an interesting game element. We observed that the effect diminishes when the scaling factor is set to extreme values ($200x$ in our pretests), and the miniaturization impression is replaced by the bird's eye perspective, i.e., players feel more like flying. This was a surprising observation, and might also be part of future research in that area.

\subsubsection{Embedding in Storytelling} 
A body of famous literature, such as \textit{Alice in Wonderland}~\cite{carroll2011alice} or \textit{Gulliver's Travels}~\cite{swift1995gulliver}, describes size alterations of the protagonists. VR games based on these stories are canonical examples where GulliVR navigation can be easily and meaningfully integrated into the storyline. On the other hand, our study shows that the technique also works without any story-related cues. Hence, both options are feasible, and the choice should depend on the actual game.

\subsubsection{Interplay with Other Game Mechanics}
GM provides an excellent overview of the current area, limiting certain quests where players have to search for a specific location. To overcome that limitation, discoverable locations could be hidden in GM and require players to explicitly switch to NM, e.g., a dungeon entry is visible only when the player is on the ground. An alternative is to cover potential points of interest in fog of war, which prevents players from gathering too detailed knowledge from above. Depending on the player task, it might be beneficial to explicitly deactivate GulliVR for certain key locations or having GulliVR only as a fast travel mechanism.

\subsubsection{Transformation Time}
Based on previous research, we assumed that a slow transformation between GM and NM would induce cybersickness. A small experiment with five participants confirmed that the transition should be executed in less than $0.005 * Scale_{GM}$ seconds and at constant speed. At lower speeds, severe cybersickness symptoms occur almost instantly. As a rule of thumb, we recommend keeping the transition always below one second. Instant transformation is also safe, but might result in a slight disorientation.

\subsubsection{Walking Ground}
Games usually rely on the ground relief to compute the player camera height, i.e., stepping on a virtual stone or hill increases the camera height. GulliVR demands certain attention at this point, as such implementations would cause a considerable amount of camera shaking while, e.g., walking over a forest in GM. To prevent that, we recommend smoothing approaches such as Gaussian blur or simple averaging, as outlined in \FG{fig:misc}. Roughly speaking, such methods align with having a giant foot while being in GM.

\subsubsection{Object Manipulation}
VR games often include the task of picking up and carrying an item to a destination. If the carried object remains visible to the player, we suggest that switching to GM should also proportionally enlarge that item. However, the ability to drop objects while being in GM should be considered. In our case, the dropped object remained big, and more than once, the interaction was a source of amusement during the free exploration mode. Furthermore, a decision needs to be made whether players in GM are allowed to interact with ``miniature'' objects on the ground.

\begin{figure}
\centering
\includegraphics[width=1.0\columnwidth]{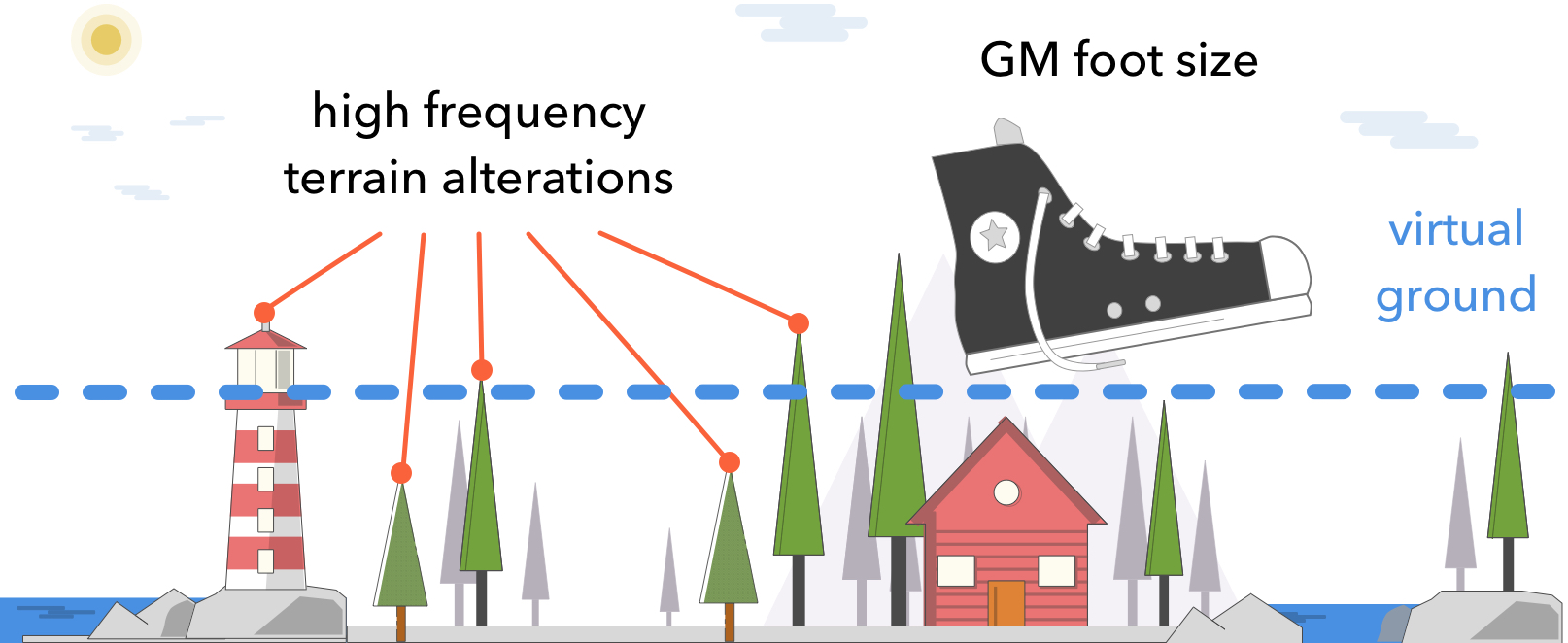}
\caption{A virtual walking ground based on the smoothed ground relief prevents camera jittering when players walk over obstacles.}
\label{fig:misc}
\end{figure}

\subsubsection{Closed, Sheltered Areas}
GulliVR works best when players do not see a ceiling above them. Paired with a fast, but not instant, transition time between GM and NM, that restriction allows us to maximize the orientation ability of players, as demonstrated in our experiment results. Although certain exceptions, e.g., going from NM to GM while being in a house in our game, work fine, we do not recommend GulliVR for games with dominant indoor scenery. Obviously, the technique is most unsuitable for narrow, multilevel closed environments.

\subsubsection{Game Genres} 
We consider 3D adventures and similar genres to be most suited for GulliVR, as such games usually revolve around emerging into a virtual world. Integrating GulliVR into fast-paced games that include enemy interactions, e.g., shooters, requires decisions regarding enemy behavior while the player is in GM. For instance, enemies could freeze, become invisible, or just be invulnerable. Furthermore, we encourage the integration of GulliVR into, e.g., RPGs, as a realistic alternative for fast travel.

\section{Conclusion and Future Work}

Implementing navigation is one of the most compelling challenges when designing a VR game. Our presented technique is a novel possibility for traversing larger distances by enlarging the players' virtual bodies on demand. In contrast to established methods such as teleportation, GulliVR emphasizes physical walking, which leads to a significantly increased presence. Furthermore, the proportionally enlarged modeled eye distance resolves the cognitive mismatch between physical movements and visual feedback, reliably obviating cybersickness. Our experiments confirmed these assumptions and promoted the suitability of our technique for various kinds of VR games and other VR applications. In addition, the paper provided a discussion on several key features such as targeting, resetting, miniaturization, and transformation time.

Our future work will investigate these features in more detail. In particular, we will explore how GulliVR can be combined with other navigation techniques to add support for traveling in sheltered environments and to allow efficient resetting mechanisms. Furthermore, our research will tackle the question whether and how the technique can be embedded into game storytelling, both in terms of players' embodiment perception and the impression of suddenly arriving in a toy world. Related to this, it might be interesting to investigate whether users should be able to adjust their giant size scale. We also suggest an evaluation of the strengths and weaknesses of GulliVR in different game genres to create a comprehensive guideline for VR researchers and developers.

\balance{}

\bibliographystyle{SIGCHI-Reference-Format}
\bibliography{gullivr}


\begin{thebibliography}{00}


\ifx \showCODEN    \undefined \def \showCODEN     #1{\unskip}     \fi
\ifx \showDOI      \undefined \def \showDOI       #1{{\tt DOI:}\penalty0{#1}\ }
  \fi
\ifx \showISBNx    \undefined \def \showISBNx     #1{\unskip}     \fi
\ifx \showISBNxiii \undefined \def \showISBNxiii  #1{\unskip}     \fi
\ifx \showISSN     \undefined \def \showISSN      #1{\unskip}     \fi
\ifx \showLCCN     \undefined \def \showLCCN      #1{\unskip}     \fi
\ifx \shownote     \undefined \def \shownote      #1{#1}          \fi
\ifx \showarticletitle \undefined \def \showarticletitle #1{#1}   \fi
\ifx \showURL      \undefined \def \showURL       #1{#1}          \fi

\bibitem{argelaguet2016giant}
{Ferran Argelaguet} {and} {Morgant Maignant}. 2016.
\newblock \showarticletitle{GiAnt: stereoscopic-compliant multi-scale
  navigation in VEs}. In {\em Proceedings of the 22nd ACM Conference on Virtual
  Reality Software and Technology}. ACM, 269--277.
\newblock


\bibitem{Bernhaupt2010}
{Regina Bernhaupt}. 2010.
\newblock {\em User Experience Evaluation in Entertainment}.
\newblock Springer London, London, 3--7.
\newblock
\showISBNx{978-1-84882-963-3}
\showDOI{%
\url{http://dx.doi.org/10.1007/978-1-84882-963-3_1}}


\bibitem{best1996perceptual}
{Scot Best}. 1996.
\newblock \showarticletitle{Perceptual and oculomotor implications of
  interpupillary distance settings on a head-mounted virtual display}. In {\em
  Aerospace and Electronics Conference, 1996. NAECON 1996., Proceedings of the
  IEEE 1996 National}, Vol.~1. IEEE, 429--434.
\newblock


\bibitem{Bhandari:2017:LSW:3139131.3139133}
{Jiwan Bhandari}, {Sam Tregillus}, {and} {Eelke Folmer}. 2017.
\newblock \showarticletitle{Legomotion: Scalable Walking-based Virtual
  Locomotion}. In {\em Proceedings of the 23rd ACM Symposium on Virtual Reality
  Software and Technology} {\em (VRST '17)}. ACM, New York, NY, USA, Article
  18, 8 pages.
\newblock
\showISBNx{978-1-4503-5548-3}
\showDOI{%
\url{http://dx.doi.org/10.1145/3139131.3139133}}


\bibitem{Biocca:1995:IVR:207922.207926}
{Frank Biocca} {and} {Ben Delaney}. 1995.
\newblock \showarticletitle{Communication in the Age of Virtual Reality}.
\newblock L. Erlbaum Associates Inc., Hillsdale, NJ, USA, Chapter Immersive
  Virtual Reality Technology, 57--124.
\newblock
\showISBNx{0-8058-1550-3}
\showURL{%
\url{http://dl.acm.org/citation.cfm?id=207922.207926}}


\bibitem{bruder2009arch}
{Gerd Bruder}, {Frank Steinicke}, {and} {Klaus~H Hinrichs}. 2009.
\newblock \showarticletitle{Arch-explore: A natural user interface for
  immersive architectural walkthroughs}.
\newblock  (2009).
\newblock


\bibitem{cairns2014immersion}
{Paul Cairns}, {Anna Cox}, {and} {A~Imran Nordin}. 2014.
\newblock \showarticletitle{Immersion in digital games: review of gaming
  experience research}.
\newblock {\em Handbook of digital games\/} (2014), 337--361.
\newblock


\bibitem{carroll2011alice}
{Lewis Carroll}. 2011.
\newblock {\em Alice's adventures in wonderland}.
\newblock Broadview Press.
\newblock


\bibitem{cho2014evaluating}
{Isaac Cho}, {Jialei Li}, {and} {Zachary Wartell}. 2014.
\newblock \showarticletitle{Evaluating dynamic-adjustment of stereo view
  parameters in a multi-scale virtual environment}. In {\em 3D User Interfaces
  (3DUI), 2014 IEEE Symposium on}. IEEE, 91--98.
\newblock


\bibitem{vive}
{HTC Corporation}. 2018.
\newblock {HTC Vive}.
\newblock Website.   (2018).
\newblock
\newblock
\shownote{Retrieved March 29, 2018 from \url{https://www.vive.com/}.}


\bibitem{cutting1995perceiving}
{James~E Cutting} {and} {Peter~M Vishton}. 1995.
\newblock \showarticletitle{Perceiving layout and knowing distances: The
  integration, relative potency, and contextual use of different information
  about depth}.
\newblock In {\em Perception of space and motion}. Elsevier, 69--117.
\newblock


\bibitem{UQO.2004}
{U.~Q.O. {Cyberpsychology Lab}}. 2004.
\newblock Presence Questionnaire: Revised by the UQO Cyberpsychology Lab.
\newblock   (2004).
\newblock
\showURL{%
\url{http://w3.uqo.ca/cyberpsy/docs/qaires/pres/PQ_va.pdf}}


\bibitem{Darken:1997:OTL:263407.263550}
{Rudolph~P. Darken}, {William~R. Cockayne}, {and} {David Carmein}. 1997.
\newblock \showarticletitle{The Omni-directional Treadmill: A Locomotion Device
  for Virtual Worlds}. In {\em Proceedings of the 10th Annual ACM Symposium on
  User Interface Software and Technology} {\em (UIST '97)}. ACM, New York, NY,
  USA, 213--221.
\newblock
\showISBNx{0-89791-881-9}
\showDOI{%
\url{http://dx.doi.org/10.1145/263407.263550}}


\bibitem{7892386}
{C. Elvezio}, {M. Sukan}, {S. Feiner}, {and} {B. Tversky}. 2017.
\newblock \showarticletitle{Travel in large-scale head-worn VR: Pre-oriented
  teleportation with WIMs and previews}. In {\em 2017 IEEE Virtual Reality
  (VR)}. 475--476.
\newblock
\showISSN{2375-5334}
\showDOI{%
\url{http://dx.doi.org/10.1109/VR.2017.7892386}}


\bibitem{engel2008psychophysically}
{David Engel}, {Crist{\'o}bal Curio}, {Lili Tcheang}, {Betty Mohler}, {and}
  {Heinrich~H B{\"u}lthoff}. 2008.
\newblock \showarticletitle{A psychophysically calibrated controller for
  navigating through large environments in a limited free-walking space}. In
  {\em Proceedings of the 2008 ACM symposium on Virtual reality software and
  technology}. ACM, 157--164.
\newblock


\bibitem{fernandes2016combating}
{Ajoy~S Fernandes} {and} {Steven~K Feiner}. 2016.
\newblock \showarticletitle{Combating VR sickness through subtle dynamic
  field-of-view modification}. In {\em 3D User Interfaces (3DUI), 2016 IEEE
  Symposium on}. IEEE, 201--210.
\newblock


\bibitem{frenz2007estimation}
{Harald Frenz}, {Markus Lappe}, {Marina Kolesnik}, {and} {Thomas B{\"u}hrmann}.
  2007.
\newblock \showarticletitle{Estimation of travel distance from visual motion in
  virtual environments}.
\newblock {\em ACM Transactions on Applied Perception (TAP)\/} {4}, 1 (2007),
  3.
\newblock


\bibitem{Grechkin:2016:RDT:2931002.2931018}
{Timofey Grechkin}, {Jerald Thomas}, {Mahdi Azmandian}, {Mark Bolas}, {and}
  {Evan Suma}. 2016.
\newblock \showarticletitle{Revisiting Detection Thresholds for Redirected
  Walking: Combining Translation and Curvature Gains}. In {\em Proceedings of
  the ACM Symposium on Applied Perception} {\em (SAP '16)}. ACM, New York, NY,
  USA, 113--120.
\newblock
\showISBNx{978-1-4503-4383-1}
\showDOI{%
\url{http://dx.doi.org/10.1145/2931002.2931018}}


\bibitem{Habgood:2017:HLP:3130859.3131437}
{M.P.~Jacob Habgood}, {David Wilson}, {David Moore}, {and} {Sergio Alapont}.
  2017.
\newblock \showarticletitle{HCI Lessons From PlayStation VR}. In {\em Extended
  Abstracts Publication of the Annual Symposium on Computer-Human Interaction
  in Play} {\em (CHI PLAY '17 Extended Abstracts)}. ACM, New York, NY, USA,
  125--135.
\newblock
\showISBNx{978-1-4503-5111-9}
\showDOI{%
\url{http://dx.doi.org/10.1145/3130859.3131437}}


\bibitem{heeter1992being}
{Carrie Heeter}. 1992.
\newblock \showarticletitle{Being there: The subjective experience of
  presence}.
\newblock {\em Presence: Teleoperators \& Virtual Environments\/} {1}, 2
  (1992), 262--271.
\newblock


\bibitem{hettinger1992visually}
{Lawrence~J Hettinger} {and} {Gary~E Riccio}. 1992.
\newblock \showarticletitle{Visually induced motion sickness in virtual
  environments}.
\newblock {\em Presence: Teleoperators \& Virtual Environments\/} {1}, 3
  (1992), 306--310.
\newblock


\bibitem{ijsselsteijn2007characterising}
{Wijnand IJsselsteijn}, {Yvonne De~Kort}, {Karolien Poels}, {Audrius
  Jurgelionis}, {and} {Francesco Bellotti}. 2007.
\newblock \showarticletitle{Characterising and measuring user experiences in
  digital games}. In {\em International conference on advances in computer
  entertainment technology}, Vol.~2. 27.
\newblock


\bibitem{ijsselsteijn2008measuring}
{Wijnand IJsselsteijn}, {Wouter Van Den~Hoogen}, {Christoph Klimmt}, {Yvonne
  De~Kort}, {Craig Lindley}, {Klaus Mathiak}, {Karolien Poels}, {Niklas
  Ravaja}, {Marko Turpeinen}, {and} {Peter Vorderer}. 2008.
\newblock \showarticletitle{Measuring the experience of digital game
  enjoyment}. In {\em Proceedings of Measuring Behavior}. Noldus Information
  Tecnology Wageningen, Netherlands, 88--89.
\newblock


\bibitem{IJsselsteijn.2013}
{W.~A. IJsselsteijn}, {Y.~A.W. de Kort}, {and} {K. Poels}. 2013.
\newblock The Game Experience Questionnaire: Development of a self-report
  measure to assess the psychological impact of digital games. Manuscript in
  Preparation.  (2013).
\newblock


\bibitem{IJsselsteijn}
{Wijnand~A. IJsselsteijn}, {Huib de Ridder}, {Jonathan Freeman}, {and}
  {Steve~E. Avons}. 2000.
\newblock Presence: concept, determinants, and measurement.
\newblock   (2000).
\newblock
\showDOI{%
\url{http://dx.doi.org/10.1117/12.387188}}


\bibitem{omni}
{Virtuix Inc.} 2018.
\newblock {Virtuix Omni}.
\newblock Website.   (2018).
\newblock
\newblock
\shownote{Retrieved March 29, 2018 from \url{http://www.virtuix.com/}.}


\bibitem{interrante2007seven}
{Victoria Interrante}, {Brian Ries}, {and} {Lee Anderson}. 2007.
\newblock \showarticletitle{Seven league boots: A new metaphor for augmented
  locomotion through moderately large scale immersive virtual environments}. In
  {\em 3D User Interfaces, 2007. 3DUI'07. IEEE Symposium on}. IEEE.
\newblock


\bibitem{iskenderova2017drunk}
{Aliya Iskenderova}, {Florian Weidner}, {and} {Wolfgang Broll}. 2017.
\newblock \showarticletitle{Drunk Virtual Reality Gaming: Exploring the
  Influence of Alcohol on Cybersickness}. In {\em Proceedings of the Annual
  Symposium on Computer-Human Interaction in Play}. ACM, 561--572.
\newblock


\bibitem{jun2015big}
{Eunice Jun}, {Jeanine~K Stefanucci}, {Sarah~H Creem-Regehr}, {Michael~N
  Geuss}, {and} {William~B Thompson}. 2015.
\newblock \showarticletitle{Big foot: Using the size of a virtual foot to scale
  gap width}.
\newblock {\em ACM Transactions on Applied Perception (TAP)\/} {12}, 4 (2015),
  16.
\newblock


\bibitem{kennedy1993simulator}
{Robert~S Kennedy}, {Norman~E Lane}, {Kevin~S Berbaum}, {and} {Michael~G
  Lilienthal}. 1993.
\newblock \showarticletitle{Simulator sickness questionnaire: An enhanced
  method for quantifying simulator sickness}.
\newblock {\em The international journal of aviation psychology\/} {3}, 3
  (1993), 203--220.
\newblock


\bibitem{kennedy1989simulator}
{Robert~S Kennedy}, {Michael~G Lilienthal}, {Kevin~S Berbaum}, {DR Baltzley},
  {and} {ME McCauley}. 1989.
\newblock \showarticletitle{Simulator sickness in US Navy flight simulators.}
\newblock {\em Aviation, Space, and Environmental Medicine\/} {60}, 1 (1989),
  10--16.
\newblock


\bibitem{kokkinara2015effects}
{Elena Kokkinara}, {Mel Slater}, {and} {Joan L{\'o}pez-Moliner}. 2015.
\newblock \showarticletitle{The effects of visuomotor calibration to the
  perceived space and body, through embodiment in immersive virtual reality}.
\newblock {\em ACM Transactions on Applied Perception (TAP)\/} {13}, 1 (2015),
  3.
\newblock


\bibitem{kolasinski1995simulator}
{Eugenia~M Kolasinski}. 1995.
\newblock {\em Simulator Sickness in Virtual Environments.}
\newblock {T}echnical {R}eport. Army Research Inst. for the Behavioral and
  Social Sciences, Alexandria, VA.
\newblock


\bibitem{kopper2006design}
{Regis Kopper}, {Tao Ni}, {Doug~A Bowman}, {and} {Marcio Pinho}. 2006.
\newblock \showarticletitle{Design and evaluation of navigation techniques for
  multiscale virtual environments}. In {\em Virtual Reality Conference, 2006}.
  Ieee, 175--182.
\newblock


\bibitem{lambooij2009visual}
{Marc Lambooij}, {Marten Fortuin}, {Ingrid Heynderickx}, {and} {Wijnand
  IJsselsteijn}. 2009.
\newblock \showarticletitle{Visual discomfort and visual fatigue of
  stereoscopic displays: A review}.
\newblock {\em Journal of Imaging Science and Technology\/} {53}, 3 (2009),
  30201--1.
\newblock


\bibitem{langbehn2016subliminal}
{Eike Langbehn}, {Gerd Bruder}, {and} {Frank Steinicke}. 2016.
\newblock \showarticletitle{Subliminal reorientation and repositioning in
  virtual reality during eye blinks}. In {\em Proceedings of the 2016 Symposium
  on Spatial User Interaction}. ACM, 213--213.
\newblock


\bibitem{laviola2000discussion}
{Joseph~J LaViola~Jr}. 2000.
\newblock \showarticletitle{A discussion of cybersickness in virtual
  environments}.
\newblock {\em ACM SIGCHI Bulletin\/} {32}, 1 (2000), 47--56.
\newblock


\bibitem{laviola2001hands}
{Joseph~J LaViola~Jr}, {Daniel~Acevedo Feliz}, {Daniel~F Keefe}, {and}
  {Robert~C Zeleznik}. 2001.
\newblock \showarticletitle{Hands-free multi-scale navigation in virtual
  environments}. In {\em Proceedings of the 2001 symposium on Interactive 3D
  graphics}. ACM, 9--15.
\newblock


\bibitem{lin2002effects}
{JJ-W Lin}, {Henry Been-Lirn Duh}, {Donald~E Parker}, {Habib Abi-Rached}, {and}
  {Thomas~A Furness}. 2002.
\newblock \showarticletitle{Effects of field of view on presence, enjoyment,
  memory, and simulator sickness in a virtual environment}. In {\em Virtual
  Reality, 2002. Proceedings. IEEE}. IEEE, 164--171.
\newblock


\bibitem{lombard1997heart}
{Matthew Lombard} {and} {Theresa Ditton}. 1997.
\newblock \showarticletitle{At the heart of it all: The concept of presence}.
\newblock {\em Journal of Computer-Mediated Communication\/} {3}, 2 (1997),
  0--0.
\newblock


\bibitem{medeiros2016effects}
{Daniel Medeiros}, {Eduardo Cordeiro}, {Daniel Mendes}, {Maur{\'\i}cio Sousa},
  {Alberto Raposo}, {Alfredo Ferreira}, {and} {Joaquim Jorge}. 2016.
\newblock \showarticletitle{Effects of speed and transitions on target-based
  travel techniques}. In {\em Proceedings of the 22nd ACM Conference on Virtual
  Reality Software and Technology}. ACM, 327--328.
\newblock


\bibitem{money1970motion}
{K~E Money}. 1970.
\newblock \showarticletitle{Motion sickness.}
\newblock {\em Physiological Reviews\/} {50}, 1 (1970), 1--39.
\newblock


\bibitem{nguyen2011effects}
{Tien~Dat Nguyen}, {Christine~J Ziemer}, {Timofey Grechkin}, {Benjamin Chihak},
  {Jodie~M Plumert}, {James~F Cremer}, {and} {Joseph~K Kearney}. 2011.
\newblock \showarticletitle{Effects of scale change on distance perception in
  virtual environments}.
\newblock {\em ACM Transactions on Applied Perception (TAP)\/} {8}, 4 (2011),
  26.
\newblock


\bibitem{ogawa2017distortion}
{Nami Ogawa}, {Takuji Narumi}, {and} {Michitaka Hirose}. 2017.
\newblock \showarticletitle{Distortion in perceived size and body-based scaling
  in virtual environments}. In {\em Proceedings of the 8th Augmented Human
  International Conference}. ACM, 35.
\newblock


\bibitem{ohyama2007autonomic}
{Seizo Ohyama}, {Suetaka Nishiike}, {Hiroshi Watanabe}, {Katsunori Matsuoka},
  {Hironori Akizuki}, {Noriaki Takeda}, {and} {Tamotsu Harada}. 2007.
\newblock \showarticletitle{Autonomic responses during motion sickness induced
  by virtual reality}.
\newblock {\em Auris Nasus Larynx\/} {34}, 3 (2007), 303--306.
\newblock


\bibitem{Poels:2007:ALF:1328202.1328218}
{Karolien Poels}, {Yvonne de Kort}, {and} {Wijnand IJsselsteijn}. 2007.
\newblock \showarticletitle{"It is Always a Lot of Fun!": Exploring Dimensions
  of Digital Game Experience Using Focus Group Methodology}. In {\em
  Proceedings of the 2007 Conference on Future Play} {\em (Future Play '07)}.
  ACM, New York, NY, USA, 83--89.
\newblock
\showISBNx{978-1-59593-943-2}
\showDOI{%
\url{http://dx.doi.org/10.1145/1328202.1328218}}


\bibitem{poupyrev1996go}
{Ivan Poupyrev}, {Mark Billinghurst}, {Suzanne Weghorst}, {and} {Tadao
  Ichikawa}. 1996.
\newblock \showarticletitle{The go-go interaction technique: non-linear mapping
  for direct manipulation in VR}. In {\em Proceedings of the 9th annual ACM
  symposium on User interface software and technology}. ACM, 79--80.
\newblock


\bibitem{razzaque2005redirected}
{Sharif Razzaque}. 2005.
\newblock {\em Redirected walking}.
\newblock University of North Carolina at Chapel Hill.
\newblock


\bibitem{razzaque2001redirected}
{Sharif Razzaque}, {Zachariah Kohn}, {and} {Mary~C Whitton}. 2001.
\newblock \showarticletitle{Redirected walking}. In {\em Proceedings of
  EUROGRAPHICS}, Vol.~9. Citeseer, 105--106.
\newblock


\bibitem{reason1975motion}
{James~T Reason} {and} {Joseph~John Brand}. 1975.
\newblock {\em Motion sickness.}
\newblock Academic press.
\newblock


\bibitem{renner2015influence}
{Rebekka~S Renner}, {Erik Steindecker}, {Mathias M{\"u}Ller}, {Boris~M
  Velichkovsky}, {Ralph Stelzer}, {Sebastian Pannasch}, {and} {Jens~R Helmert}.
  2015.
\newblock \showarticletitle{The influence of the stereo base on blind and
  sighted reaches in a virtual environment}.
\newblock {\em ACM Transactions on Applied Perception (TAP)\/} {12}, 2 (2015),
  7.
\newblock


\bibitem{renner2013perception}
{Rebekka~S Renner}, {Boris~M Velichkovsky}, {and} {Jens~R Helmert}. 2013.
\newblock \showarticletitle{The perception of egocentric distances in virtual
  environments-a review}.
\newblock {\em ACM Computing Surveys (CSUR)\/} {46}, 2 (2013), 23.
\newblock


\bibitem{ruddle2009benefits}
{Roy~A Ruddle} {and} {Simon Lessels}. 2009.
\newblock \showarticletitle{The benefits of using a walking interface to
  navigate virtual environments}.
\newblock {\em ACM Transactions on Computer-Human Interaction (TOCHI)\/} {16},
  1 (2009), 5.
\newblock


\bibitem{ruddle2011walking}
{Roy~A Ruddle}, {Ekaterina Volkova}, {and} {Heinrich~H B{\"u}lthoff}. 2011.
\newblock \showarticletitle{Walking improves your cognitive map in environments
  that are large-scale and large in extent}.
\newblock {\em ACM Transactions on Computer-Human Interaction (TOCHI)\/} {18},
  2 (2011), 10.
\newblock


\bibitem{Schubert.1999b}
{T.~W. Schubert}, {Frank Friedmann}, {and} {H.~T. Regenbrecht}. 1999.
\newblock \showarticletitle{Decomposing the sense of presence: Factor analytic
  insights}. In {\em 2nd international workshop on presence}, Vol. 1999.
\newblock


\bibitem{sherman2002understanding}
{William~R Sherman} {and} {Alan~B Craig}. 2002.
\newblock {\em Understanding virtual reality: Interface, application, and
  design}.
\newblock Elsevier.
\newblock


\bibitem{slater2003note}
{Mel Slater}. 2003.
\newblock \showarticletitle{A note on presence terminology}.
\newblock {\em Presence connect\/} {3}, 3 (2003), 1--5.
\newblock


\bibitem{slater1995virtual}
{Mel Slater}, {Anthony Steed}, {and} {Martin Usoh}. 1995a.
\newblock \showarticletitle{The virtual treadmill: A naturalistic metaphor for
  navigation in immersive virtual environments}.
\newblock In {\em Virtual Environments' 95}. Springer, 135--148.
\newblock


\bibitem{slater1995taking}
{Mel Slater}, {Martin Usoh}, {and} {Anthony Steed}. 1995b.
\newblock \showarticletitle{Taking steps: the influence of a walking technique
  on presence in virtual reality}.
\newblock {\em ACM Transactions on Computer-Human Interaction (TOCHI)\/} {2}, 3
  (1995), 201--219.
\newblock


\bibitem{stanney1997cybersickness}
{Kay~M Stanney}, {Robert~S Kennedy}, {and} {Julie~M Drexler}. 1997.
\newblock \showarticletitle{Cybersickness is not simulator sickness}. In {\em
  Proceedings of the Human Factors and Ergonomics Society annual meeting},
  Vol.~41. SAGE Publications Sage CA: Los Angeles, CA, 1138--1142.
\newblock


\bibitem{steinicke2007hybrid}
{Frank Steinicke}, {Gerd Bruder}, {and} {Klaus Hinrichs}. 2007.
\newblock \showarticletitle{Hybrid traveling in fully-immersive large-scale
  geographic environments}. In {\em Proceedings of the 2007 ACM symposium on
  Virtual reality software and technology}. ACM, 229--230.
\newblock


\bibitem{stoakley1995virtual}
{Richard Stoakley}, {Matthew~J Conway}, {and} {Randy Pausch}. 1995.
\newblock \showarticletitle{Virtual reality on a WIM: interactive worlds in
  miniature}. In {\em Proceedings of the SIGCHI conference on Human factors in
  computing systems}. ACM Press/Addison-Wesley Publishing Co., 265--272.
\newblock


\bibitem{suma2010evaluation}
{Evan Suma}, {Samantha Finkelstein}, {Myra Reid}, {Sabarish Babu}, {Amy
  Ulinski}, {and} {Larry~F Hodges}. 2010.
\newblock \showarticletitle{Evaluation of the cognitive effects of travel
  technique in complex real and virtual environments}.
\newblock {\em IEEE Transactions on Visualization and Computer Graphics\/}
  {16}, 4 (2010), 690--702.
\newblock


\bibitem{swift1995gulliver}
{Jonathan Swift}. 1995.
\newblock \showarticletitle{Gulliver's travels}.
\newblock In {\em Gulliver's Travels}. Springer, 27--266.
\newblock


\bibitem{unity}
{Unity Technologies}. 2018.
\newblock {Unity}.
\newblock Website.   (2018).
\newblock
\newblock
\shownote{Retrieved March 29, 2018 from \url{https://unity3d.com/}.}


\bibitem{tregillus2016vr}
{Sam Tregillus} {and} {Eelke Folmer}. 2016.
\newblock \showarticletitle{Vr-step: Walking-in-place using inertial sensing
  for hands free navigation in mobile vr environments}. In {\em Proceedings of
  the 2016 CHI Conference on Human Factors in Computing Systems}. ACM,
  1250--1255.
\newblock


\bibitem{usoh1999walking}
{Martin Usoh}, {Kevin Arthur}, {Mary~C Whitton}, {Rui Bastos}, {Anthony Steed},
  {Mel Slater}, {and} {Frederick~P Brooks~Jr}. 1999.
\newblock \showarticletitle{Walking> walking-in-place> flying, in virtual
  environments}. In {\em Proceedings of the 26th annual conference on Computer
  graphics and interactive techniques}. ACM Press/Addison-Wesley Publishing
  Co., 359--364.
\newblock


\bibitem{valkov2010traveling}
{Dimitar Valkov}, {Frank Steinicke}, {Gerd Bruder}, {and} {Klaus~H Hinrichs}.
  2010.
\newblock \showarticletitle{Traveling in 3d virtual environments with foot
  gestures and a multi-touch enabled wim}. In {\em Proceedings of virtual
  reality international conference (VRIC 2010)}. 171--180.
\newblock


\bibitem{van2011being}
{Bj{\"o}rn van~der Hoort}, {Arvid Guterstam}, {and} {H~Henrik Ehrsson}. 2011.
\newblock \showarticletitle{Being Barbie: the size of one's own body determines
  the perceived size of the world}.
\newblock {\em PloS one\/} {6}, 5 (2011), e20195.
\newblock


\bibitem{von2016cyber}
{Sebastian Von~Mammen}, {Andreas Knote}, {and} {Sarah Edenhofer}. 2016.
\newblock \showarticletitle{Cyber sick but still having fun}. In {\em
  Proceedings of the 22nd ACM Conference on Virtual Reality Software and
  Technology}. ACM, 325--326.
\newblock


\bibitem{ware1998dynamic}
{Colin Ware}, {Cyril Gobrecht}, {and} {Mark~Andrew Paton}. 1998.
\newblock \showarticletitle{Dynamic adjustment of stereo display parameters}.
\newblock {\em IEEE transactions on systems, man, and cybernetics-part A:
  systems and humans\/} {28}, 1 (1998), 56--65.
\newblock


\bibitem{wartell1999analytic}
{Zachary~Justin Wartell}, {Larry~F Hodges}, {and} {William Ribarsky}. 1999.
\newblock {\em The analytic distortion induced by false-eye separation in
  head-tracked stereoscopic displays}.
\newblock {T}echnical {R}eport. Georgia Institute of Technology.
\newblock


\bibitem{wingrave2006overcoming}
{Chadwick~A Wingrave}, {Yonca Haciahmetoglu}, {and} {Doug~A Bowman}. 2006.
\newblock \showarticletitle{Overcoming world in miniature limitations by a
  scaled and scrolling WIM}. In {\em 3D User Interfaces, 2006. 3DUI 2006. IEEE
  Symposium on}. IEEE, 11--16.
\newblock


\bibitem{Witmer.1998}
{Bob~G. Witmer} {and} {Michael~J. Singer}. 1998.
\newblock \showarticletitle{Measuring presence in virtual environments: A
  presence questionnaire}.
\newblock {\em Presence\/} {7}, 3 (1998), 225--240.
\newblock


\bibitem{yao2014oculus}
{Richard Yao}, {Tom Heath}, {Aaron Davies}, {Tom Forsyth}, {Nate Mitchell},
  {and} {Perry Hoberman}. 2014.
\newblock \showarticletitle{Oculus vr best practices guide}.
\newblock {\em Oculus VR\/} (2014), 27--39.
\newblock


\end{thebibliography}

\end{document}
